\documentclass[]{IEEEtran}
\usepackage{listings}
\usepackage{xcolor}
\usepackage{graphicx}
\usepackage{amsmath}
\usepackage{newfloat}
\usepackage{listings}
\usepackage{algorithm}
\usepackage{algorithmic}
\usepackage{caption}
\usepackage{tcolorbox}
\tcbuselibrary{listingsutf8}

\newtcolorbox{highlightbox}{
  colback=white,
  colframe=black,
  boxrule=1.5pt,
  arc=1pt,
  left=1pt,
  right=1pt,
  top=1pt,
  bottom=1pt,
}

\usepackage{graphicx}
\begin{document}
\title{Mind the Gap: Revealing Inconsistencies Across Heterogeneous AI Accelerators}
%
%

\author{
    Elliott Wen\IEEEauthorrefmark{1},
    Sean Ma\IEEEauthorrefmark{1},
    Evan Tempero\IEEEauthorrefmark{1},
    Bruce Sham\IEEEauthorrefmark{1},
    Yousong Song\IEEEauthorrefmark{1},
    Hong Jia\IEEEauthorrefmark{1},
    Daniel Luo\IEEEauthorrefmark{2},
    Jiayi Hua\IEEEauthorrefmark{2},
    Jens Dietrich\IEEEauthorrefmark{3},
    Kaiqi Zhao\IEEEauthorrefmark{4}
    Jiaxing Shen\IEEEauthorrefmark{5}\\,
    \IEEEauthorrefmark{1}The University of Auckland, New Zealand\\
    \IEEEauthorrefmark{2}Hong Kong Polytechnic University, Hong Kong\\
    \IEEEauthorrefmark{3}Victoria University of Wellington, New Zealand\\
    \IEEEauthorrefmark{4}Harbin Institute of Technology, China\\
    \IEEEauthorrefmark{5}Lingnan University, Hong Kong\\
    Email: elliott.wen, sean.ma, e.temper, bruce.sham, y.song, jia.hong@auckland.ac.nz,\\
daniel.luo, jiayi.hua@polyu.edu.hk, \\
    jens.dietrich@vuw.ac.nz, zhaokaiqi@hit.edu.cn, jiaxingshen@ln.edu.hk
}

\maketitle              
\begin{abstract}
While NVIDIA remains the dominant provider of AI accelerators within cloud data center, emerging vendors such as AMD, Intel, Mac, and Huawei offer cost-effective alternatives with claims of compatibility and performance. This paper presents the first empirical study investigating divergence in machine learning model across heterogeneous AI accelerators. Utilizing an automated pipeline, we synthesize over 100,000 variant models derived from 4,000 real-world models and execute them across five different enterprise-grade accelerators.  
Our findings suggest that newer AI platforms from Mac and Huawei support at least 17\% fewer operators than NVIDIA. These platforms also exhibit a higher rate of output discrepancies (exceeding 5\%), which stem from differences in operator implementations, handling of exceptional numerical values, and instruction scheduling. They are also more susceptible to failures during model compilation-based acceleration, and in some cases, the compiled models produce outputs that differ noticeably from those generated using the standard execution mode.
In addition, we identify 7 implementation flaws in PyTorch and 40 platform-specific issues across vendors. These results underscore the challenges of achieving consistent machine learning behavior in an increasingly diverse hardware ecosystem.

\end{abstract}

\section{Introduction}

Machine learning (ML) is a rapidly expanding field with growing influence across various disciplines. To meet the computational demands of ML models training and inference, many cloud service providers now offer access to hardware such as graphics processing units and specialized AI accelerators. Although NVIDIA-based solutions are most widely adopted, other platforms such as AMD, Intel, Mac and Huawei are gaining traction. They claim to offer sufficient computational performance for general machine learning tasks, while at a more competitive price point. In addition, they emphasize support for existing ML  frameworks, particularly PyTorch, one of the most widely used frameworks in academia and industry~\cite{imambi2021pytorch}.

As hardware platforms become increasingly diverse, concerns about their consistency and reliability are growing, particularly for developers seeking to avoid vendor lock-in and adopt more cost-effective cloud AI hardware services.
In this paper, we present, to the best of our knowledge, the first empirical study assessing divergence in ML model execution across heterogeneous AI accelerator. We aim to answer the following research question: \emph{How consistent are the outputs of ML models across different hardware platforms, and what are the potential sources of inconsistency?} To address this, we employ a differential testing approach; we execute 100,000 synthesized models with the same randomly generated inputs across various hardware platforms. We then investigate significant discrepancies to uncover potential issues in the underlying hardware or software stack.

\begin{figure*}[t]
    \centering
    \includegraphics[width=1\linewidth]{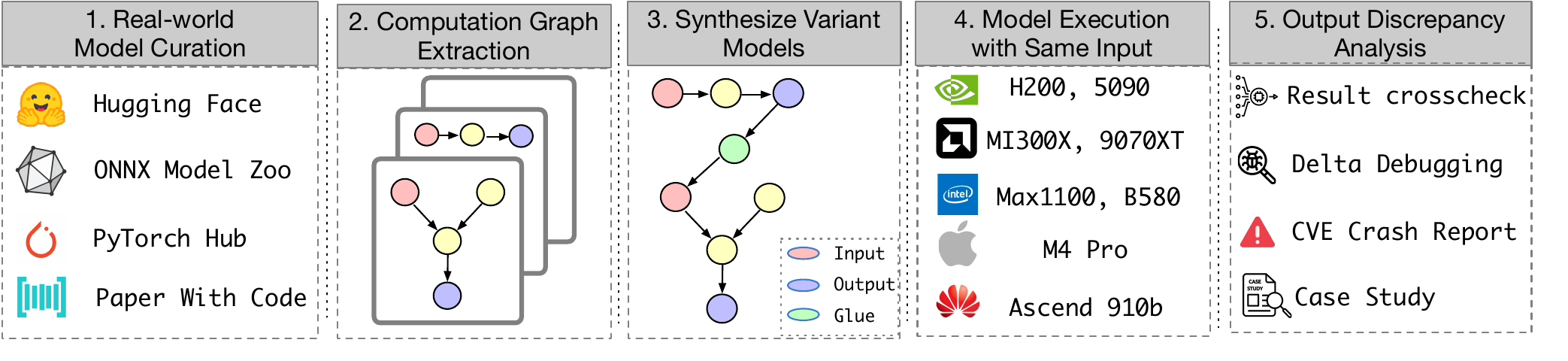}
    \caption{Our pipeline to uncover the behavior inconsistency across different hardware platforms}
    \label{fig:pytorch-pipeline}
\end{figure*}

To support this investigation, we implement an automated pipeline, as illustrated in Figure~\ref{fig:pytorch-pipeline}. 
The pipeline begins with corpus curation, where we crawl approximately 4,000 real-world models from various machine learning domains. These curated models are then converted into computational graphs, which serve as input to a model synthesizer. The synthesizer generates a large number of variant models by iteratively merging random subgraphs from the corpus and applying mutations to individual nodes, inspired by prior work~\cite{green2022graphfuzz}. The variant models are executed across different hardware platforms, and any output discrepancies are recorded. We then analyze these discrepancies to pinpoint potential sources of divergence.

We conduct our experiments on five enterprise-grade AI accelerator hardware commonly accessible through cloud, including the \texttt{NVIDIA H200}, \texttt{AMD MI300X}, \texttt{Intel Max 1100}, \texttt{Huawei Ascend 910B}, and \texttt{Mac M4 Pro}. We unveil the following key research findings:
\begin{enumerate}

\item Relatively new platforms, such as Mac and Huawei, exhibit lower model compatibility due to unsupported data types or unimplemented operations. Our analysis reveals that Mac and Huawei support 34\% and 17\% fewer operators than NVIDIA, respectively. As such, code modifications are often necessary. Most missing operators on Huawei and Mac involve quantized/sparse operations, attention, NLP embeddings, and network training.

\item Using NVIDIA as the baseline, Intel and AMD generally produce more consistent results, with output agreement rates of 99.6\% and 99.8\%, respectively. In contrast, the Ascend and Mac platforms show significantly lower consistency, with agreement rates dropping to 95\% and 86\%. These discrepancies primarily stem from flawed operator implementations, differences in handling exceptional values, variations in instruction scheduling, and instances of undefined behavior.

\item Some platforms exhibit reduced reliability when PyTorch's compilation-based acceleration features are enabled. On Mac and Huawei devices, approximately 2\% of models fail to compile, and about 1.1\% yield results that differ from those produced by the default execution mode.

\end{enumerate}

\section{Background and Related Work}
\begin{figure}[t]
    \centering
    \includegraphics[width=1\linewidth]{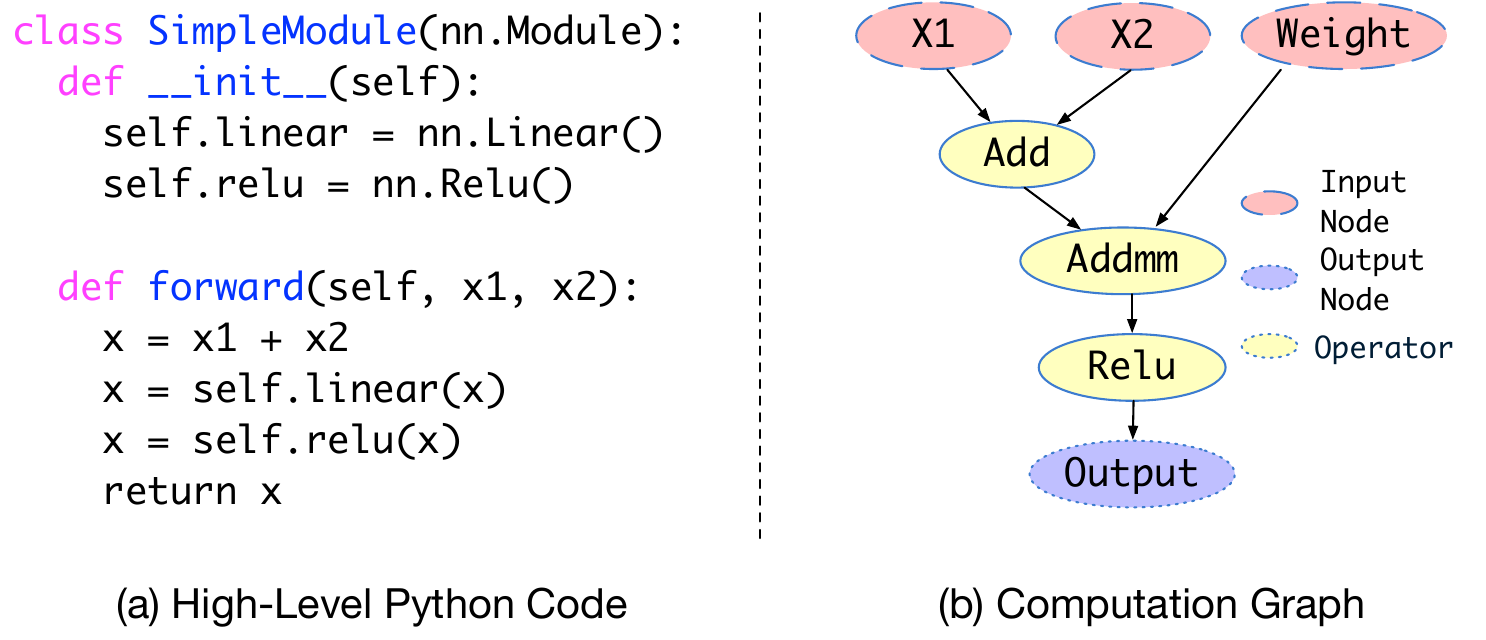}
    \caption{PyTorch Model Representation: From High-Level Python Code to Computation Graph}
    \label{fig:pytorch-internal}
\end{figure}

We open-source our testing pipeline together with our model corpus. Using this pipeline, we uncover 7 implementation flaws in the PyTorch framework, such as memory access violations and program hangs caused by infinite optimization loops in the just-in-time compiler. We also identified behavioral inconsistencies or implementation flaws across different platforms: 1 on NVIDIA, 4 on AMD, 3 on Intel, 11 on Mac, and 13 on Huawei.

\textbf{ML Model and Computation Graph:} 
To aid reader comprehension of the rest of this paper, we first provide a brief overview of ML frameworks. We focus on PyTorch due to its broad adoption in both academic research and industry applications. Moreover, Pytorch is the only framework that consistently receives active development and optimization support from all five hardware vendors discussed in this paper.

In Pytorch, models are defined in Python by subclassing the \verb|nn.Module| class. The model's  layers are specified within the \verb|init| method, while the forward computation logic is implemented in the \verb|forward| method. Once defined, input data can be passed through the model to generate predictions. Figure~\ref{fig:pytorch-internal}(a) illustrates a simple model that takes two input tensors, $x1$ and $x2$, adds them together, and passes the result through a linear layer followed by a ReLU activation function. Under the hood, these high-level Python tensor operations are dispatched to PyTorch's low-level C++ backend, known as \texttt{ATen}. Each operation is mapped to a corresponding ATen operator. In our example, the tensor addition, linear transformation, and ReLU activation are translated into \verb|aten::add|, \verb|aten::addmm|, and \verb|aten::relu|, respectively. These ATen operators form a directed acyclic computation graph as shown in Figure~\ref{fig:pytorch-internal}(b), where each operation corresponds to a node. An edge is created between two nodes if the output of one operation is used as the input to another. The input and output tensors of the graph can be treated as special types of nodes, representing the entry and exit points of the computation. Internally, ATen operators are further redispatched to hardware-specific implementations, such as CUDA kernels \texttt{relu\_cuda} and \texttt{addmm\_cuda} to leverage hardware acceleration. 

\noindent\textbf{Framework Bug Detection:} ML frameworks can exhibit various types of bugs, such as incorrect operator implementations or memory safety violations~\cite{morovati2024bug}. Existing approaches for bug detection typically involve feeding data into curated or  synthesized models to identify framework-level errors or inconsistencies.
For instance,
CRADLE~\cite{pham2019cradle} targets cross-framework inconsistencies by comparing outputs from a small set of real-world models across multiple frameworks. Rather than relying on real-world models, subsequent works such as AUDEE~\cite{guo2020audee} and Luo et al.~\cite{luo2021graph} generate models and inputs using genetic algorithms and graph-theoretic techniques. Building on this direction, LEMON~\cite{alzamily2020lemon}, Muffin~\cite{gu2022muffin}, and EAGLE~\cite{wang2022eagle} introduce various model mutation strategies to expand the pool of test models.  NNSmith~\cite{liu2023nnsmith} proposes advanced abstract operator modeling to further enhance model diversity.

More recently, researchers have begun exploring differential testing at the operator level instead of the model level. For example, FreeFuzz~\cite{wei2022free}, SkipFuzz~\cite{kang2022skipfuzz} and IvySyn~\cite{christou2023ivysyn} apply random mutations, such as changes to data types and values, to individual operator parameters to uncover potential bugs. 
DeepREL~\cite{deng2022fuzzing} and TENSORSCOPE~\cite{deng2023differential} identify related APIs within a library and corresponding APIs across different libraries. They then leverage test inputs from other APIs to generate inputs for the target APIs.
DocTer~\cite{xie2022docter} and ACETest~\cite{shi2023acetest} extract and analyze input constraints from API documentation and C++ implementation to generate more valid and targeted parameter inputs. 


Our testing pipeline draw inspiration from prior research and introduces a practical extension by combining graph synthesizing techniques with operator-level input variation methods to explore a broader diversity of execution behaviors.


\noindent\textbf{Our Contributions:} 
Unlike prior work that primarily investigates software-level bugs and inconsistencies within ML frameworks, our study explores a complementary dimension: behavioral divergence across heterogeneous AI accelerators. 
Our findings on platform compatibility and reliability have practical implications for developers aiming to avoid vendor lock-in and adopt more cost-effective hardware alternatives.

\section{Pipeline Implementation}
In this section, we provide implementation details for our testing pipeline. 

\subsection{Dataset Curation}
To synthesize variant models for testing, we first require a model corpus. We construct this corpus from a large and diverse collection of real-world machine learning models from multiple domains.

We implement a web crawler to collect models from popular machine learning distribution platforms such as Hugging Face\footnote{https://huggingface.co/} and TorchHub\footnote{https://pytorch.org/hub/}. These platforms offer convenient access to a wide range of pre-trained models in various domains and enable deployment with minimal Python code.
Our crawler successfully collected approximately 170,000 PyTorch models from 50 tasks domains along with their associated metadata. 
While the total number of models is substantial, many share the same underlying network architecture. The primary differences often lie in their learned weights, as they are just fine-tuned on different datasets or downstream tasks. To eliminate duplicate architectures, we examine the accompanying \verb|config.json| files, which specify key architectural parameters such as the model class name and the number of layers. We also analyze each model's architecture-specific base class (e.g., \verb|BertModel| or \verb|GPT2Model|) to assess structural equivalence. Eventually, we retain a final set of 3,711  models with unique architectures.

 
In addition, we manually curate models from the \textit{Papers with Code} platform\footnote{https://github.com/paperswithcode/paperswithcode-data}, which connects ML publications with their corresponding code implementations and datasets. 
Unlike the previously mentioned sources, Papers with Code is academically oriented and exhibits greater variation in code structure and optimization techniques. We retrieve the platform's database as of May 2025 and identify 71,232 PyTorch models across 45 categories. From these, we apply random sampling to select a total of 300 repositories.




\subsection{Computation Graph Extraction}
The next step is capture the computational graph of each model in the corpus. We achieve this by instrumenting the Pytorch's ATen dispatcher functions, such as \verb|kernel.call| and \verb|kernel.callBoxed|, to capture detailed operator invocation records.
For each invocation, we record its input parameters and return values. Most parameters are simple constants, such as scalar numbers or boolean values. If a parameter is a tensor, we additionally record its identifier along with metadata such as shape, data type, and storage device. 
The recorded traces reveal how tensors are produced and consumed across operator invocations. Using this information, we can reconstruct the corresponding computational graph through a straightforward one-to-one mapping. This approach records computation graph transparently without altering the original model and supports both forward and backward computation graph. In contrast, PyTorch's in-built graph tracing frameworks, such as GraphFX or JIT tracing, require modifying model code and do not capture backward computation graph.

\subsection{Variant Model Generation} 
We adopt the graph merging technique introduced in GraphFuzz~\cite{green2022graphfuzz} to generate variant models. This method has demonstrated high network structural diversity and code coverage.
Algorithm~\ref{alg:variant-model-generation} outlines the procedure. The process begins with an empty graph $G$, and iteratively integrates subgraphs extracted from our computational graph corpus until a predefined node threshold $T$ is reached. As $T$ increases, the architectural diversity expands, which also leads to increased model execution and compilation time. Based on extensive preliminary experiments, we set $T = 1{,}000$ nodes to achieve a trade-off between them.

A key implementation detail involves resolving mismatches in tensor dimensions between the output nodes of the existing graph and the input nodes of the new components. This is achieved by inserting additional glue nodes. If the output tensor contains more elements than required, we first apply the \verb|aten::flatten| operator to convert the output tensor into a one-dimensional form, followed by \verb|aten::slice| to remove the excess elements, and finally use \verb|aten::reshape| to match the desired input shape. Conversely, if the output tensor contains fewer elements than needed, we employ the \verb|aten::pad| operator to insert additional constants before reshaping the tensor.

To further enhance the diversity of our models, we also employ a well-tested operator mutation strategy from prior work~\cite{wei2022free}. Specifically, with a predefined probability (e.g., 0.25), we replace an operator with a syntactically similar alternative. For instance, an out-of-place operator such as \verb|aten::add| may be substituted with its in-place counterpart \verb|aten::add_|, which stores the result directly in the input tensor rather than creating a new one. Activation functions can also be replaced with alternatives, such as substituting \verb|aten::relu| with \verb|aten::hardtanh|. Additionally, functionally equivalent operations with different formulations may be exchanged, for example, replacing \verb|aten::add| with \verb|aten::addcdiv| with appropriate scaling and divisor tensors.

\begin{algorithm}[t]
\caption{Variant Model Generation Workflow}
\label{alg:variant-model-generation}
\begin{algorithmic}[1]
\STATE Initialize an empty graph $G$
\WHILE{number of nodes in $G$ $\leq$ threshold $T$}
    \STATE Sample a computational graph $C$ from the corpus
    \STATE Extract a random connected subgraph $S$ from $C$
    \FORALL{incoming edges in $S$ without a predecessor node}
        \STATE Insert a new input node as its predecessor
    \ENDFOR
    \FORALL{outgoing edges in $S$ without a subsequent node}
        \STATE Insert a new output node as its successor
    \ENDFOR
    \STATE Merge $S$ into $G$ by randomly connecting output nodes of $G$ to input nodes of $S$
\ENDWHILE
\end{algorithmic}
\end{algorithm}

\subsection{Model Execution}

We evaluate each synthesized model across three execution modes in PyTorch.
\begin{enumerate}
\item{\textbf{Default Mode (i.e., Eager Mode)}}: In this mode, computation operations are executed immediately as they are encountered in the Python code (i.e., without optimization).
Eager mode is the most widely used execution mode in PyTorch. Our analysis of all GitHub repositories listed on the \textit{Papers with Code} platform reveals that approximately 97\% of them solely use Eager mode.
To execute in this mode, we first deserialize the synthesized graphs using the \verb|torch.jit.load| API, which reconstructs the corresponding high-level Python statements. These statements are then executed directly using the \verb|eval| API.

\item{\textbf{Just-In-Time (JIT) Compilation}}: In JIT mode, the model initially runs in eager mode for several iterations to gather runtime information such as input tensor shapes and data types. This collected information is then used to transform the graph through a series of simple-yet-effective optimization passes, such as constant folding and dead code elimination, to enhance execution efficiency. 
We can enable JIT profiling and compilation using the \verb|torch._C._jit_set_profiling_mode| API. 

\item{\textbf{TorchDynamo Compilation}}: PyTorch recently introduces a new execution mode called \textit{TorchDynamo}. 
In this mode, PyTorch hands off the computation graph to vendor-specific backends (e.g., IPEX for Intel) that compile it into highly optimized machine code. 
To run in this mode, we pass the deserialized Python module directly to the \verb|torch.compile| API. 
\end{enumerate}

We generate a set of random inputs to evaluate each model. We adjust the value ranges according to the required input types as guided by prior work~\cite{wang2024d}.  
For example, floating-point tensors (e.g., \texttt{torch.double}) are populated with values sampled uniformly from the range \([0, 1)\).  
Special consideration is given to the \texttt{torch.int64} data type, which is frequently used as an index parameter in many aten operators. Excessively large indices can easily lead to out-of-bounds accesses. To suppress the error, we adopt a heuristic range of \([0, 4]\), informed by our preliminary experiments.
Following previous work~\cite{xie2022docter}, we also adopt a set of input constraints to ensure semantic correctness; for example, the offset array in \verb|aten::embedding_bags| must begin with 0. We also disable the autocast feature to ensure all the test cases to run in the same precision mode.



\subsection{Output Comparison and Analysis Discrepancy}
Following prior work~\cite{wang2022eagle}, we adopt the inconsistency detection formula in the TensorFlow and PyTorch test suites. Specifically, two outputs \( O_{p1} \) and \( O_{p2} \) from different platforms are considered equivalent between outputs if the condition $
|O_{p1} - O_{p2}| \leq \text{atol} + \text{rtol} \cdot |O_{p2}|$ holds element-wise, where \( \text{atol} = 5 \times 10^{-4} \) is the absolute tolerance and \( \text{rtol} = 1 \times 10^{-4} \) is the relative tolerance. 
Both thresholds are used jointly to account for absolute and relative differences in the outputs. 


Once output discrepancies are detected across platforms, we aim to trace them back to the specific locations in the model that triggered the differences to facilitate manual analysis.
To achieve this, we construct a data flow graph of the model to capture the dependency chain between output tensors and the operations that produce them. 
Beginning with the earliest operator in the dependency chain, we perform a layer-by-layer comparison of the outputs generated on each platform. This iterative process continues until we isolate the first operator whose output diverges. As part of this process, we adopt the comparison metrics from CRADLE~\cite{pham2019cradle}. Specifically, we compute the Mean Absolute Distance (MAD) between each pair of corresponding outputs in the dependency chain across platforms. We then calculate the rate of change for each output and cluster inconsistencies based on the operator that exhibits the highest rate of change.



\section{Evaluation}
We conduct our evaluations on a bare-metal machine equipped with a 240-core 5th Gen Intel Xeon Platinum 8580 processor and 1.97 TiB of DDR5 memory. It is installed with four enterprise-grade AI accelerator hardware platforms, including the \texttt{NVIDIA H200}, \texttt{AMD MI300X}, \texttt{Intel MAX 1100}, and \texttt{Huawei Ascend 910B}. To include the Mac platform in our study, we additionally use a \texttt{Mac M4} Pro desktop machine. We use the NVIDIA H200 as our baseline, given its widespread deployment in AI production environments.

\subsection{Discrepency in Model Execution across Platforms} 
In our experiment, we synthesized 100,000 variant models from our corpus. In our setup, the model generation process takes approximately 17 minutes. 
The execution times are approximately 289 hours on the \texttt{NVIDIA H200}, 297 hours on the \texttt{Intel MAX 1100}, 320 hours on the \texttt{AMD MI300X}, 277 hours on the \texttt{Mac M4 Pro}, and 412 hours on the \texttt{Huawei Ascend 910B}.
Huawei exhibits slightly longer execution times, likely due to its unoptimized PyTorch extension. Specifically, it introduces an initialization overhead (about 15 seconds in our setup) each run to query the available hardware operators. 

\begin{figure}[t]
    \centering
    \includegraphics[width=0.9\linewidth]{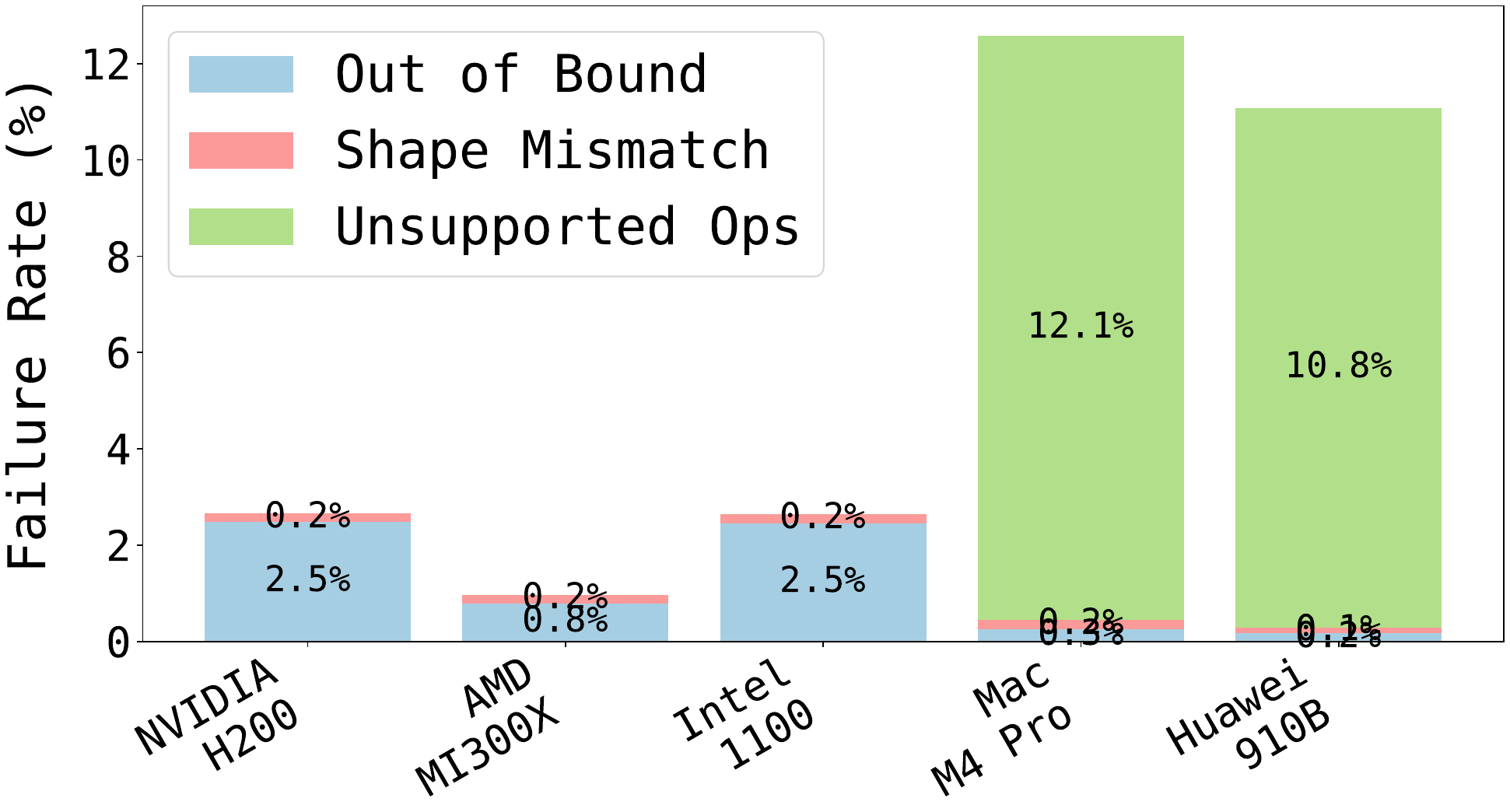}
    \caption{Model Execution Failures in Default Execution Mode}
    \label{fig:pytorch-execution}
\end{figure}

We record the execution behavior of each model in default execution mode and visualize the number of failures and their root causes across different platforms in Figure~\ref{fig:pytorch-execution}. Each model is evaluated using the latest version of its corresponding runtime environment.

\noindent\textbf{Baseline:}
For our baseline NVIDIA platform, 
the most common type of failure is out-of-bounds errors, which are expected given that the models are tested with randomly generated inputs, including index tensors. Although these index tensors are generated within a restricted range, out-of-bounds accesses may still occur, for example, when the tensor being indexed is also small.
The other type of error is shape mismatch. These errors arise because certain operators, such as \texttt{aten::where} or \texttt{aten::select}, determine the shape of their output tensors based on the values of the input tensors. Due to input randomness, the resulting shapes of intermediate tensors may differ from the shapes expected by subsequent operations, thus leading to execution failures.
Nevertheless, these two categories of errors together account for only about 2.5\% of total model executions on our baseline platform.
Following prior work~\cite{wei2022free}, we also examine the successfully executed models and analyze the types of operators they contain. Our analysis reveals that these models cover 397 out of the 488 operators available in PyTorch 2.7.1. This indicates that our synthesized models achieve reasonable operator coverage.

\noindent\textbf{Unsupported Operations on Huawei and Mac:} Another notable phenomenon is that the Mac and Huawei platforms exhibit a significant number of execution failures due to unsupported operation errors. They account for more than 10\% of all models.
Our pipeline detects 47 unimplemented operators on Mac and 34 on Huawei. It is important to note that these numbers likely underestimate the count of unsupported operators, as some operators silently fall back to CPU implementations without warning. Such fallbacks often lead to significant performance degradation. 

To gain a deeper understanding on the unimplemented operators, Specifically, we examine the code graph to identify operators that ultimately invoke hardware command interfaces (e.g., \verb|EXEC_NPU_CMD| on Huawei). This indicates that these operators can benefit from hardware acceleration. We observe that our NVIDIA baseline provides hardware acceleration for 488 operators, with the AMD platform offering an identical level of support. Intel supports slightly fewer operators, totaling 449. Huawei’s platform falls further behind and supports 407 operators, which is 17.5 percent fewer than NVIDIA. The Mac platform offers the least support, with hardware acceleration available for just 332 operators. This is 34.0 percent fewer than NVIDIA. 
The missing operators on these platforms fall into the following categories:

\noindent{1. Quantized Inference:} Huawei and Mac offer very limited support for quantized inference operators, such as \verb|aten::_weight_int4pack_mm|. These operators enable matrix multiplication using compressed data formats to reduce memory consumption and bandwidth usage.

\noindent{2. Sparse Operations:} Both platforms lack support for sparse tensor computation like \verb|aten::_sparse_semi_structured_linear|. These routines exploit structural sparsity in model weights to decrease memory usage and computation time by skipping zero elements.

\noindent{3. Attention Mechanism:} Huawei and Mac do not provide specialized attention kernels, such as \verb|aten::flash_attention|. These kernels fuse multiple steps of the attention computation into a single optimized operation for higher computation speed.

\noindent{4. NLP Embeddings:} Both platforms lack support for embedding operations like \verb|aten::_embedding_bag_backward|. They are crucial for computing gradients in embedding bag layers commonly found in NLP models.

\noindent{5. Fused Training Support:} Fused implementations of the Adam optimizer update (e.g., \verb|aten::fused_adam|) and dropout steps (e.g., \verb|aten::fused_dropout|) are missing on Huawei and Mac, which potentially degrades the training efficiency.

\noindent{6. Advanced Linear Algebra:} Huawei and Mac lack support for advanced linear algebra routines, such as \verb|aten::triangular_solve| and \verb|aten::inalg_matrix_exp|, which are essential in scientific computing, control theory, and differentiable physics.


What makes matters worse is that most operators on the Mac and Huawei platforms cannot handle 64-bit floating-point input tensors. Additionally, some operators on the Huawei platform do not support 64-bit integers as the data type for index tensors. The Mac backend consistently raises a runtime error when encountering a 64-bit floating-point value. In contrast, the behavior on the Huawei Ascend platform is less predictable. For certain operators (e.g., \texttt{aten::to}), the data may be silently cast to a lower-precision type, while some operators (e.g., \verb|aten::huber_loss_backward|) trigger a runtime error. We also observed that operators on Mac and Huawei platforms sometimes require input tensors to be in a contiguous memory format. Otherwise, an exception may be raised.





\noindent\textbf{Missing Bounds Checking on AMD:} Another interesting observation is that AMD platforms report significantly fewer out-of-bounds (OOB) errors compared to the baseline and Intel platforms (approximately 1,700 fewer). Upon examining the models that trigger OOB on AMD, we find that all of them are reported by the baseline and Intel platforms. Conversely, many additional models flagged as OOB by the baseline execute without error on AMD and produce outputs that may appear correct to a human observer.
Moreover, OOB errors on the baseline and Intel platforms typically raise runtime exceptions. These exceptions include clear stack traces, which make it easier to identify the offending operators. In contrast, the AMD platform often returns a generic error message such as `memory access violation' and crashes the PyTorch program. This provides little diagnostic insight to users.

\begin{highlightbox}
Finding 1:  Mac and Huawei platforms lack support for certain operators and data types. As a result, developers may need to refactor their codebases, potentially at the cost of reduced performance and precision. 
\end{highlightbox}

\begin{highlightbox}
Finding 2: Most missing operators on Huawei and Mac involve quantized/sparse operations, attention, NLP embeddings, and Adam training, indicating reduced efficiency for common deep learning workloads.
\end{highlightbox}
\begin{highlightbox}
Finding 3: AMD lacks a robust bounds-checking mechanism, which hinders the early detection of model bugs and introduces potential security vulnerabilities.
\end{highlightbox}

\subsection{Inconsistency in Model Outputs and Case Studies}
\begin{figure}[t]
    \centering
    \includegraphics[width=0.9\linewidth]{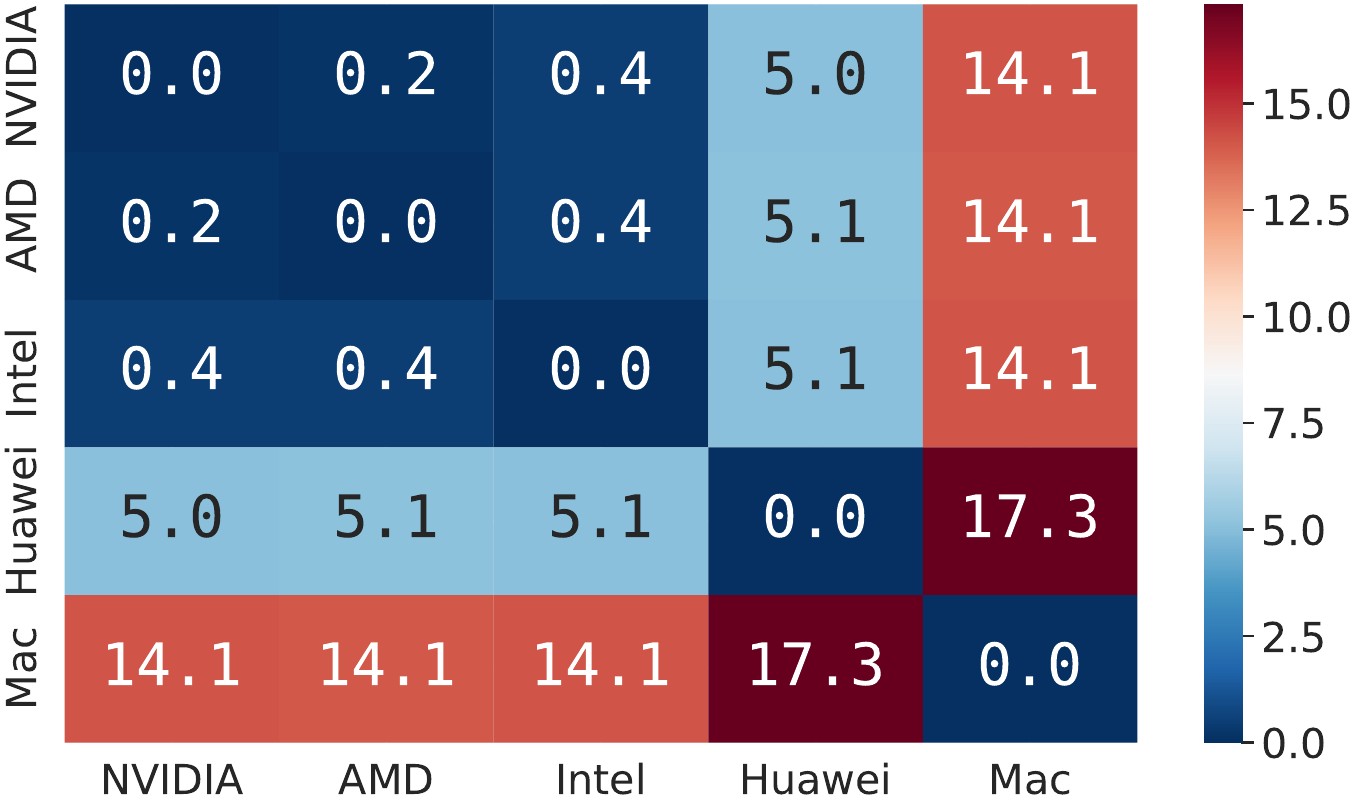}
 
    \caption{Model Output Consistency Across Platforms}
    \label{fig:hardwareconsistency}
\end{figure}

We compare model outputs across different hardware platforms. We use only 87,840 models from the previous section that run successfully on all the platforms.
Figure~\ref{fig:hardwareconsistency} illustrates the observed result discrepancies.  
Notably, AMD demonstrates the lowest deviation from the baseline, affecting only 0.2\% of the evaluated models. Intel shows a slightly higher discrepancy rate at 0.4\%. Ascend exhibits a moderate deviation and impacts 5.1\% of the models, while Mac has the highest discrepancy rate and affects 14.1\% of the models.

To better understand the nature of these inconsistencies, we performed a manual analysis on a sample of inconsistent cases and present a case study. We categorize the cases as follows.

\noindent \textbf{1) Operator Implementation:} Output inconsistencies may stem from flawed operator implementations. We identified 13 faulty operators on the Huawei platform that produce incorrect results, data types, or tensor shapes.
For example, the operator \verb|max_pool1d_with_indices|, when provided with a 64-bit integer index tensor, unexpectedly returns a tensor with elements of an unsigned 8-bit data type. The operator \verb|grid_sampler_2d_backward| appears to produce incorrect outputs in the zero padding mode, whereas all other platforms yield consistent results. Additionally, the operator \verb|prelu_kernel_backward| is expected to return two tensors with matching dimensions; however, this condition does not hold on the Huawei platform. Most of the identified operators have the suffix `backward', further suggesting that the Huawei platform may be less suitable for training tasks.
We also observe 11 faulty operators on the Mac platform. For instance, the widely used operator \texttt{aten::pad}, which adds extra elements around the edges of a tensor, may fail to function correctly; when the padding size exceeds a threshold (e.g., $65,\!536$ for a 1D tensor), the operator begins to overwrite existing elements with corrupted data. This issue is particularly concerning given the operator's widespread usage.



\noindent \textbf{2) Exceptional Number Handling and Propagation:} Operators on different platforms may handle exceptional numerical values differently. For instance, when encountering numerically unstable inputs, the operators \texttt{aten::remainder} and \texttt{aten::convolution} produce \texttt{Inf} on AMD and Intel, whereas on NVIDIA they return \texttt{NaN}. We also observe variations in how exceptional values (e.g., \texttt{Inf} and \texttt{NaN}) propagate through the computation graph. Ideally, such values should be consistently propagated across operators to aid developers in identifying numerical issues. However, AMD and Mac do not always follow this practice. For example, on AMD, the \verb|aten::batch_norm| operator appears to replace \texttt{NaN} elements with interpolated values from nearby data. Mac is even less consistent: common operators like \verb|aten::reshape| have been observed to silently convert \texttt{NaN} values into zeros.

\noindent \textbf{3) Instruction Scheduling:} 
Some operators can exhibit nondeterministic behavior, in other words, even with identical inputs, their outputs may vary slightly across different platforms or runs. This variability often stems from the parallel execution of these operators; inconsistencies can arise from how the compiler or runtime system schedules hardware instructions. For example, we observe that the operator \verb|aten::max_unpool2d|, when given an index tensor containing repeated elements, behaves deterministically on NVIDIA hardware. Our 10,000 repeated runs all produce identical outputs. In contrast, on AMD hardware, the same operator exhibits nondeterministic behavior under the same conditions. The 10,000 runs yields three different outputs, each occurring at varying ratios


\noindent \textbf{4) Undefined Behavior}: 
Different computing platforms employ distinct strategies for handling undefined operations. One notable example is positive integer division by zero. NVIDIA returns \texttt{4,294,967,295}, the maximum value of a 32-bit unsigned integer. On macOS, the result is \texttt{0}, while AMD interestingly returns the original value incremented by one. Another example involves converting floating-point infinity to a 64-bit integer. On NVIDIA, this conversion yields the largest possible 64-bit integer, whereas AMD returns \texttt{-4,294,967,296}.

\begin{highlightbox}
Finding 4: Intel and AMD generally produce results more consistent with the baseline than Mac and Huawei.
\end{highlightbox}

\begin{highlightbox}
Finding 5: Platform inconsistency is mainly attributed to flawed operator implementations, exceptional value handling, instruction scheduling, and undefined behavior.
\end{highlightbox}

\begin{table*}[t]
\centering
\caption{Summary of model compilation issues across platforms}
\label{tab:model-issues}
\resizebox{0.8\linewidth}{!}{%
\begin{tabular}{|l|c|c|c|c|c|}
\hline
\textbf{Issue Type} & \textbf{Baseline} & \textbf{AMD} & \textbf{Intel} & \textbf{Mac} & \textbf{Huawei} \\ \hline
\multicolumn{6}{|l|}{\textbf{JIT}} \\ \hline
Stalled Compilation         & 9 & 9 & 3  & 7 & 12 \\ \hline
Heap Corruption
& 2 & 3 & 7 & 3  & 27 \\ \hline
\textbf{Total}              & \textbf{11 ($<$0.1\%)} & \textbf{12 ($<$0.1\%)} & \textbf{10 ($<$0.1\%)} & \textbf{10 ($<$0.1\%)} & \textbf{39 ($<$0.1\%)} \\ \hline
\multicolumn{6}{|l|}{\textbf{TorchDynamo}} \\ \hline
Stalled Compilation         & 0  & 0  & 0  & 0  & 16 \\ \hline
Unsupported Ops             & 22  & 22  & 79 & 1719  & 1484 \\ \hline
Heap Corruption                       & 4 & 3 & 8 & 3  & 45 \\ \hline
Codegen Error      & 3  & 3  & 0  & 0  & 0  \\ \hline

\textbf{Total}              & \textbf{29 ($<$0.1\%)} & \textbf{28 ($<$0.1\%)} & \textbf{87 (0.1\%)} & \textbf{1722 (2.2\%)} & \textbf{1529 (1.9\%)} \\ \hline
\end{tabular}
}
\end{table*}

\subsection{Analysis of Model Compilation Issues} 
We re-execute the models discussed in the previous section using PyTorch's compilation modes, namely JIT and TorchDynamo. This evaluation includes only $77,291$ models  that yield consistent outputs across all platforms under the default execution mode.
We expect the output of the compiled models to remain consistent with that produced by the default execution mode, aside from very minor numerical discrepancies. Any substantial deviation may indicate an unreliable implementation within the vendor-specific compiler.

\begin{figure}[t]
    \centering
    \includegraphics[width=0.9\linewidth]{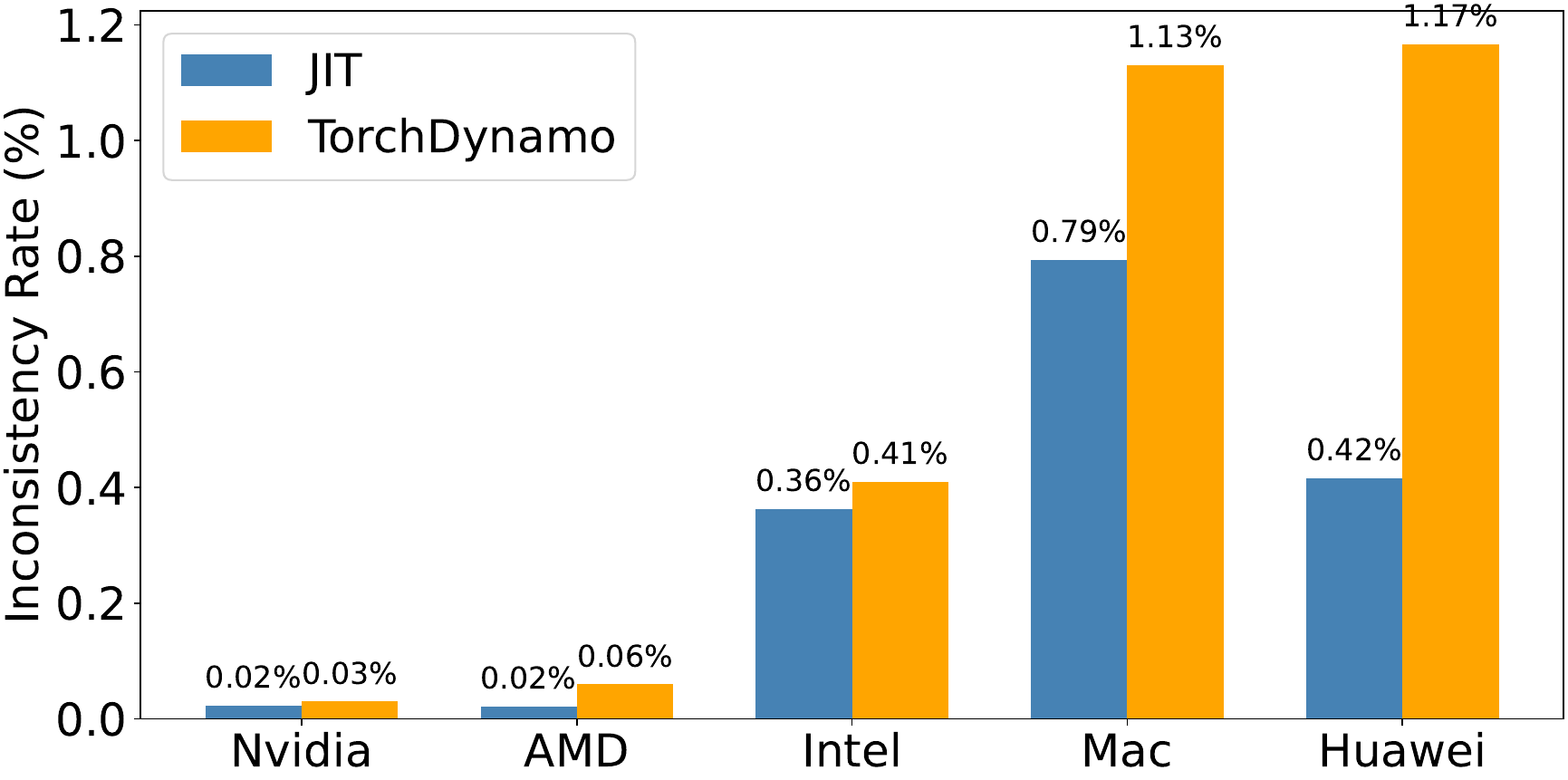}
    \caption{Result Consistency Between Compilation Mode and Eager Mode}
    \label{fig:result-consistency}
\end{figure}

Figure~\ref{fig:result-consistency} presents the count of inconsistencies between the results obtained using the two compilation modes and those produced by the default execution mode. Overall, JIT compilation appears to be more reliable than TorchDynamo, likely due to its simpler optimization passes. Among all platforms, NVIDIA and AMD demonstrate the fewest inconsistencies.
We manually examine these cases and find that the inconsistencies primarily involve transitions between \texttt{NaN} and \texttt{Inf}. This may be attributed to the reordering of numerically unstable operations. Mac and Huawei exhibit the highest levels of inconsistency, approximately 1.1\%. Our analysis indicates that most discrepancies involve significant differences in numerical values. This suggests potential implementation errors within the vendor-specific compiler pipeline.

We encountered a number of failures during the compilation process and a summary of the findings is provided in Table~\ref{tab:model-issues}. One key observation is that platforms generally exhibit better compatibility (i.e., fewer failures) with PyTorch's JIT compilation mode compared to TorchDynamo. 
In addition, compatibility also varies across platforms, with some experiencing significantly higher rates of compilation failures, for example, 1.9\% on Huawei compared to 0.1\% on the Baseline. We manually analyze these issues and categorize them as follows:

\noindent{\textbf{1) Stalled Compilation:}} We observed a JIT compiler hanging issue affecting almost every platforms. Our analysis attributes this behavior to a flaw in PyTorch's JIT optimization passes. Specifically, one pass attempts to simplify if/else predicates statements to reduce control flow, In some cases, later passes undo earlier simplifications, causing a loop that prevents the compiler making any progress. The issue appears less severe on Intel platforms, possibly because Intel provides custom optimization passes that avoid some of this problematic behavior. 
In TorchDynamo mode, we also observed several instances of the compiler hanging, but only on the Huawei platform. Our preliminary analysis suggests that this may also be caused by flawed optimization passes.

\noindent{\textbf{2) Unsupported Ops}:}
TorchDynamo compilation encounters failures due to unsupported operations. In our evaluation, the most robust platform support was observed on NVIDIA and AMD GPUs, where all 22 failures were attributed to a small subset of operators lacking support for complex data types. In contrast, Intel platforms exhibited significantly more issues, with 172 model failures primarily caused by the lack of support for 64-bit floating-point data types in certain operators.  Mac and Huawei platforms reported approximately 80 times more unsupported operation errors than NVIDIA, suggesting that their compiler infrastructures are still maturing.

\noindent\textbf{3) Heap Corruption:} 
We observed crash instances across nearly all platforms. Manual analysis of the stack traces revealed that these crashes were primarily caused by heap corruption and segmentation faults. We identified that two of the issues originate from the PyTorch's JIT optimization pass implementation, which lacks proper bounds checking. The remaining issues stem from vendor-specific compilers. Most platforms reported fewer than 10 crash instances, whereas Huawei platforms experienced up to ten times as many. This discrepancy may be attributed to less mature compiler implementations on the latter.

\noindent\textbf{4) Code Generation Error:} 
In TorchDynamo mode, the computation graph is first converted into a C++ function and then compiled by vendor-specific backends. NVIDIA uses NVFuser for this purpose. AMD aims to maintain compatibility with CUDA and thus also adopts NVFuser. On these two platforms, 
we encountered three failures during the code translation process, specifically:
\texttt{error: duplicate parameter name}. When we examine the generated C++ code, we found that the resulting function contained multiple parameters with the same name. This leads to invalid C++ function prototypes and causes compilation failures.

\begin{highlightbox} Finding 6: Model compilation can reduce output reliability, especially on Mac and Huawei platforms. When compilation is desired to improve performance, JIT mode is recommended over TorchDynamo due to higher reliability.
\end{highlightbox}

\begin{highlightbox}
Finding 7: Platform compatibility with PyTorch's compilation modes varies considerably. Huawei and macOS show notably higher rates of compilation failures compared to other platforms.
\end{highlightbox}

\subsection{Threats to Validity}
Our current evaluation focuses on single-device execution. We plan to extend our analysis to distributed settings, where multiple AI accelerator cards operate in parallel. This may reveal additional inconsistencies arising from the complexities of hardware communication and synchronization. Additionally, our experiments are conducted using the latest available versions of runtime environments, compilers, and drivers. While this ensures relevance to current deployments, it may overlook issues present in older or less frequently updated systems. Evaluating a broader range of software versions could reveal version-specific bugs and provide a more comprehensive view of platform stability.
Finally, our study focuses solely on the latest generation of hardware. Including consumer-grade or earlier generations  could reveal hardware-specific behaviors and inconsistencies even within the same vendor. Additional research efforts are warranted to investigate this further.




\section{Conclusion}
In this paper, we explore the consistency of machine learning model outputs across five major AI accelerator platforms in cloud data centers. 
Our findings indicate that newer AI platforms from Mac and Huawei support at least 17\% fewer operators compared to NVIDIA, particularly those essential for training. These platforms also show a higher rate of output discrepancies (over 5\%), which can be attributed to differences in operator implementations, the handling of exceptional numerical values, and instruction scheduling. In addition, they are more prone to failures during model compilation based acceleration. These results highlight the difficulty of ensuring consistent machine learning behavior across an increasingly diverse range of hardware platforms.

\bibliographystyle{splncs04}
\bibliography{aaai2026}

\end{document}